\documentclass[12pt,preprint]{aastex}

\newcommand{\HD}{HD~98800~B}

\slugcomment{To appear in the Astrophysical Journal}

\shorttitle{Circumbinary disk of HD~98800~B}
\shortauthors{Akeson et al.}

\begin{document}

\title{The circumbinary disk of HD~98800~B: Evidence for disk warping }


\author{R. L. Akeson\altaffilmark{1}, W.K.M. Rice\altaffilmark{2},
A.F. Boden\altaffilmark{1}, A.I. Sargent\altaffilmark{3}, 
J.M. Carpenter\altaffilmark{3}, G. Bryden\altaffilmark{4}
}


\altaffiltext{1}{Michelson Science Center, California Institute of Technology, Pasadena, CA, 91125}
\altaffiltext{2}{Scottish Universities Physics Alliance (SUPA), Institute for Astronomy, University of Edinburgh, Blackford Hill,
Edinburgh, EH9 3HJ, Scotland}
\altaffiltext{3}{Dept. of Astronomy, California Institute of Technology, Pasadena CA 91125}
\altaffiltext{4}{Jet Propulsion Laboratory, California Institute of Technology, Pasadena, CA, 91108}

\begin{abstract}

The quadruple young stellar system HD~98800 consists of two
spectroscopic binary pairs with a circumbinary disk around the B
component.  Recent work by Boden and collaborators using infrared
interferometry and radial velocity data resulted in a determination of
the physical orbit for \HD.  We use the resulting inclination of the
binary and the measured extinction toward the B component stars to
constrain the distribution of circumbinary material.  Although a
standard optically and geometrically thick disk model can reproduce
the spectral energy distribution, it can not account for the observed
extinction if the binary and the disk are co-planar.  We next
constructed a dynamical model to investigate the influence of the A
component, which is not in the Ba-Bb orbital plane, on the B disk. We
find that these interactions have a substantial impact on the
inclination of the B circumbinary disk with respect to the Ba-Bb
orbital plane.  The resulting warp would be sufficient to place
material into the line of sight and the non-coplanar disk orientation
may also cause the upper layers of the disk to intersect the line of
sight if the disk is geometrically thick.  These simulations also
support that the dynamics of the Ba-Bb orbit clear the inner region
to a radius of $\sim$3~AU.  We then discuss whether the somewhat
unusual properties of the \HD\ disk are consistent with material
remnant from the star formation process or with more recent creation
by collisions from larger bodies.

\end{abstract}


\section{Introduction}
\label{intro}

As the majority of stars are formed in binary or higher order multiple
systems, understanding the effects of multiplicity is essential to
understanding star formation.  HD 98800 is a member of the TW Hydra
association, whose members have distances from 40 to 100 pc
\citep[{\it Hipparcos};][]{per97}, and is a quadruple system comprising
two spectroscopic binaries \citep{tor95}, HD 98800 A and B, both with
K star primaries.  The AB pair has a separation of 0\farcs{8},
corresponding to 34 AU and a period of 300-430 years \citep{tok99}.
Age estimates for the TW Hya association range from 5 to 20 Myr
\citep[and references therein]{sod98} and
several members including HD 98800 \citep{zuc93}, have
significant infrared excess emission.

\citet{geh99}, \citet{koe00} and \citet{pra01} resolved the AB system
excess and determined that the mid-infrared excess arises from
material around the B spectroscopic binary, the northern component.
The excess is detected only at wavelengths longer than 8 microns and
is well fit by a 150~K blackbody \citep{low99,koe00,pra01}.  Recent
{\it Spitzer} IRS observations by \citet{fur07} show a small excess
starting at 5.5 $\mu$m with the majority of the excess at wavelengths of
8 $\mu$m and longer.  At this temperature the material is several AU
from the stars and must therefore be circumbinary given the semi-major
axis of 0.98~AU 
\citep{bod05}.  HD~98800 shows no sign of active accretion as
judged by the H$\alpha$ line width of 0.19 Angstroms \citep{dun97} or
X-ray emission \citep{kas04}.  This is in contrast with two of the
other association members, TW Hya and Hen 3-600~A, which have observed 
accretion rates only 1-2 orders of magnitude lower than those in 1 Myr-old
T Tauri stars \citep{muz00}.

Circumbinary disks are not uncommon around closely spaced (separation
less than a few AU) young stellar binaries \citep{jen97}.  However, at
an age of 10 Myr, 80-90\% of young stars have dispersed their
primordial disk as traced by near and mid-infrared excess
\citep{hil07}.  The population of main sequence stars with an infrared
excess (the so-called debris or secondary disk systems) discovered by
IRAS \citep{aum84} has been greatly expanded by the sensitivity of
{\it Spitzer}.  Recent {\it Spitzer} observations show $\sim$15\% of A
to K spectral type field dwarfs have a mid-infrared excess
(e.g. \citet{bei06,che06}), but the material in these disks is
believed to arise from collisions between larger bodies rather than
being primordial and many of these stars have an excess only
at wavelengths $> 20\mu$m, indicative of cool dust.

Determining the evolutionary state of the \HD\ circumbinary disk is
complicated by multiplicity.  Secondary disks are also observed around
main sequence binaries \citep{tri07} as they are around pre-main
sequence binaries \citep{jen97}.  Dynamical studies have shown that a
circumbinary disk will have a cleared central region which can greatly
slow or stop material from accreting onto the central star
\citep{art94,pic05}.  Thus it is not clear if the material surrounding
B is remnant from the star formation process or has been created by
collisions from larger bodies.

Recently, \citet{bod05} used measurements from the Keck
Interferometer, the {\it Hubble Space Telescope} Fine Guidance Sensors
and radial velocity data to determine the physical orbit of \HD\ and
the mass and luminosity of its components.  \citet{bod05} modeled the
SEDs of the HD 98800 B components, finding effective temperatures of
4200 $\pm$ 150 K and 4000 $\pm$ 150 K, and luminosities of 0.330 $\pm$
0.017 L$_{\odot}$ and 0.167 $\pm$ 0.038 L$_{\odot}$ for the primary
and secondary respectively.  Additionally, a small extinction of
$A_V~0.3 \pm 0.05$ was required to fit the SED of the system.
\citet{sod98} also found that a small extinction ($A_V \approx$ 0.44
mag) was necessary to explain the lithium-predicted visible
photometry.  \citet{sod98} additionally noted the significant time variability
with no obvious periodicity in the {\it Hipparcos} optical photometry,
which \citet{tok99} suggests may be due to extinction through an
edge-on disk.  In contrast, an earlier monitoring campaign by
\citet{hen94} detected a 0.07 magnitude variation at $V$ with a 14 day
period which they suggest is due to a starspot; however, these authors
were unaware of the dual spectroscopic binary composition of the
HD~98800 system.

In this paper, we use the orbital parameters and the observed SED to
constrain the physical properties of the circumstellar material around
\HD.  In \S \ref{disk}, we investigate the origin of the extinction
towards \HD\ given the inclination of the binary orbit using
geometrically thick circumstellar disk models.  In \S \ref{dynamics},
we present a dynamical simulation of the interaction between the A and
B components and the B disk.  In \S \ref{discussion}, we compare our results
to other models of the \HD\ disk and
discuss the implications of the modeling on the question of whether
the disk material is primordial or secondary and on the origin
of the extinction along the line of sight to \HD.  Our conclusions
are summarized in \S \ref{summary}.

\section{Input data}

\subsection{Photometry}

We use near-infrared through sub-millimeter photometry from the
literature \citep{pra01, syl96} to determine the flux contribution of
the disk.  In this work, we concentrate only on the continuum emission
and do not use the photometry affected by the 10~$\mu$m silicate
feature.  \citet{koe00} and \citet{pra01} discuss this spectral
feature and recent work by \citet{fur07} using the {\it Spitzer}
mid-infrared spectrum suggests the presence of amorphous carbon,
pyroxene and olivine grains.  The stellar effective temperatures,
luminosities and system distance are taken from \citet{bod05}.

\subsection{Orbital parameters} \label{orbit}

We use the relevant orbital parameters from \citet{bod05} (Table
\ref{tab:orbit}) to derive the dynamical constraints on the disk.
Circumstellar material will tend to be cleared from the inner disk region
through interactions with the central stars \citep[see
e.g.][]{art94,pic05} and this clearing increases with the orbital
eccentricity of the binary.  For \HD\, the apastron distance is 1.75
AU, making any material within this radius dynamically unstable.  From
the calculations of \citet{pic05} the circumbinary disk for a system
such as \HD\ with eccentricity of 0.78 and secondary mass fraction
$m_2/(m_1 + m_2) =0.45$ is cleared to a radius of ~3.5 times the
semi-major axis or 3.4 AU.

\begin{table}[ht]
\begin{center}
\begin{tabular}{lrr} \tableline
e &	0.7849 $\pm$ 0.0053 \\
i (deg)&	66.8 $\pm$ 3.2 \\
a (mas)&	23.3 $\pm$ 2.5 \\
System distance (pc) &	42.2 $\pm$ 4.7 \\ \tableline
& Ba & Bb \\ \tableline
Mass (M$_{\odot}$)& 0.699 $\pm$ 0.064 &	0.582 $\pm$ 0.051 \\
T$_{eff}$ (K) &	4200 $\pm$ 150 &	4000 $\pm$ 150 \\
Luminosity (L$_{\odot}$)&0.330 $\pm$ 0.075 & 0.167 $\pm$ 0.038 \\
\tableline
\end{tabular}
\caption{Orbital and physical parameters for \HD\ from \citet{bod05}.
\label{tab:orbit}
}
\end{center}
\end{table}

The outer radius was derived using similar constraints from the A-B
orbit parameters from Tokovinin (1999; a = 62 AU and e = 0.5; Orbit II), an
A component total mass of 1.3 M$_{\odot}$ and the circumsecondary truncation
results shown in Figure 7 of \citet{art94} for a viscosity of
10$^{-3}$.  Although the \citet{art94} models are for a
circumsecondary disk in a planar orbit and the A-B and Ba-Bb orbits
are not coplanar \citep{tok99} we use this as a guide to how any
material distributed around B would be dynamically influenced by the
A-B orbit and these inner and outer truncation radii are
confirmed for the case of \HD\ by the dynamical model
presented in \S \ref{dynamics}.

\subsection{Extinction}

Mid-infrared imaging from \citet{koe00} and \citet{pra01} indicated
the large IR excess in HD 98800 is associated with HD 98800 B, with no
significant excess emission from HD 98800 A.  Spectroscopic and
interferometric observations that resolve the flux ratios from the B
subsystem components allowed \citet{bod05} to model the B subsystem SED, and
assess the individual component temperatures and luminosities.  The
\citet{bod05} SED modeling required a small amount of extinction to match
visible HST photometry; a similar conclusion was drawn by
\citet{sod98} using visible spectroscopy.  Finally, both groups
concluded that no extinction is apparent in the SED of HD 98800 A.
The association of IR excess and extinction with B and not A suggests
that there is circumbinary material around the B subsystem.

To quantify the column density of dust which would produce $A_V~0.3
\pm 0.05$, we used the dust parameters of \cite{woo02} derived for the
HH~30 disk.  These parameters are most likely to be appropriate if the
extinction arises due to the line of sight passing through some
portion of the \HD\ circumbinary disk and assuming a similar
composition for the HH~30 and \HD\ disks.  At $V$ (0.55 $\mu$m), the dust mass
opacity from \citet{woo02} is 35 cm$^2$/gm.  For an extinction of $A_V~0.3
\pm 0.05$, this results in
a column density of $8 \times 10^{-3}$ g/cm$^2$.  For comparison, the
surface density of the minimum mass solar nebula at 1 AU is $\sim10^3$
g/cm$^2$ \citep{wei77}, thus if the extinction is due to the line
of sight passing through the disk, it
is only through the tenuous outer layers.

\section{Circumbinary disk models}
\label{disk}

Several previous groups have successfully reproduced the SED of \HD\
with a $\sim$150~K blackbody \citep{low99, koe00, pra01, fur07}.
However, none of these groups included all the dynamical constraints
($3 \lesssim R_{stable} \lesssim 10$~AU) and the unresolved mid-IR imaging of
\citet{pra01}.  Here we consider if a geometrically thick dust and gas
disk can reproduce the SED while fitting both the dynamical
constraints and intersecting the line of sight to provide the observed
extinction.

We use a Bayesian approach to characterize the disk parameters; this
approach has been used to characterize disks around young stars by
\citet{lay97} and \citet{ake02}.  We fix some of the physical
characteristics based on the assumption of the circumbinary disk being
co-planar with the B binary orbit (i.e. inclination) and other
parameters based on the physics of the input model (i.e. radial
temperature gradient).  For the remaining parameters we use the
Bayesian formalism to calculate a probability distribution for each
parameter.  This method shows how well a given parameter is
constrained and how the free parameters may be degenerate with each
other, which is a particular concern using only spatially unresolved
data.  It also facilitates the inclusion of additional observational
constraints (e.g. the dynamical truncation radii).  In our case, the
assumptions made (the inclination angle and co-planarity of the disk
and the Ba-Bb orbit) do not have associated probabilities, so the
relative likelihood a single model, given the known data set, is
proportional to $e^{-\chi^2/2}$.  This probability is calculated for
every possible set of parameter values. Then, for each parameter
value, the relative probability is the sum of all probabilities for
the range of values in all the other parameters.

\subsection{Flared accretion disk}
\label{flared}

\begin{figure}[h!]
\plotfiddle{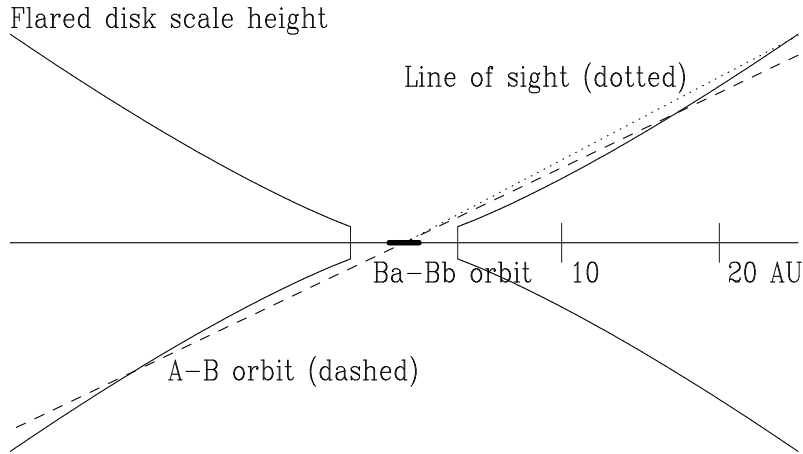}{0.1in}{270}{170}{300}{80}{0}
\caption{Illustration of the angles and scales for the HD 98800B
system.  The figure is looking edge-on to the Ba-Bb orbit (thick solid
line) with the center of mass at the origin.  The inferred
line-of-sight to Earth is shown as a dotted line and the angle with
the A-B orbit is shown as a dashed line.  The scale height for a
flared disk following the \citet{chi97} calculations (\S \ref{flared})
is also shown, with an inner radius of 3.4 AU.  The outer radius is 
deliberately unconstrained to demonstrate where the disk would intersect
the line-of-sight if the radial extent were sufficiently large.
\label{diagram}
}
\end{figure}

We first consider a disk with physical properties relevant to the
active accretion phase of younger T Tauri stars and in particular,
consider a flared disk model to investigate if the extinction can
arise from the line-of-sight intersecting the outer regions of a
flared disk.  A simple calculation shows that this is plausible.
The exact form of the disk height ($H$) as a function of radius ($r$) depends 
on the disk temperature profile and ranges from $H \propto r^{9/8}$
to $H \propto r^{9/7}$ \citep{ken87,chi97}.
Here, we use the relations derived by 
\citet{chi97} where at 
radii relevant to the \HD\ disk $H = 0.17r^{9/7}$ where $r$ is
the disk radius in AU.  Therefore the angle of incidence ($\alpha$)
for light from the central star is $\tan \alpha = H/r = 0.17r^{2/7}$.
From \citet{bod05}, the orbital inclination angle is 67\degr $\pm$ 3\degr\
and therefore $\alpha >$ 23\degr\ when $r > 25$~AU (Figure \ref{diagram}).
Although an outer radius this large is probably ruled out by the
A-B orbit truncation effects, the observed column is so small that
it may arise from material above the disk scale height.

To further investigate the possibility that the extinction arises from
the outer regions of a flared disk we parameterized the \citet{chi97}
disk and determined the disk properties as constrained by the SED
data.  The disk is comprised of an interior and a superheated surface.
For the wavelengths and radii under consideration in the \HD\ disk,
the surface layer is always optically thin and so the surface flux is simply
added to the interior flux (which may be optically thick).  To account
for a different stellar flux than in the original model, we use the
interior $T_i$ and surface $T_s$ temperature radial scalings from \citet{chi97}, but
include an single scaling factor $T_o$,
\begin{equation}
T_i = T_o (150/r^{3/7})~{\rm K}
\end{equation}
\begin{equation}
T_s = T_o (550/r^{2/5})~{\rm K},
\end{equation}
where $r$ is the radial distance in AU.
The surface density ($\Sigma$) and its radial exponent ($p$) are also allowed to vary.
The resulting model has six parameters: $T_o, \Sigma_o$ (surface density at 2 AU), $p, r_{in}, \Delta r$ (disk radial extent)
and $\beta$ (dust opacity frequency exponent) and the range of
values searched for each is given in Table \ref{tab:param}.  The inclination angle
is set at 67\degr.

\begin{table}[ht]
\begin{center}
\begin{tabular}{lcc}
Parameter & Range searched & 90\% prob. range \\ \tableline
T$_o$ & 0.35 -- 0.7 & 0.45 -- 0.6 \\
$\Sigma$ (gm/cm$^2$) & 0.02 -- 3  & 0.025 -- 0.5 \\
p & 0.5 -- 2.5 & unconstrained \\
$\Delta$r (AU) & 0.2 -- 4 & 0.8 -- 2.0  \\
r$_{in}$ (AU) & 1 -- 4 & 1.1 -- 3.6 \\
$\beta$ & 0 - 2 & 0.15 -- 0.9 \\ \tableline
\end{tabular}
\caption{Range of values searched for each parameter of the flared
disk model and the
range of values which encompasses a probability range of 90\%.
\label{tab:param}
}
\end{center}
\end{table}

\begin{figure}[!ht]
\plotfiddle{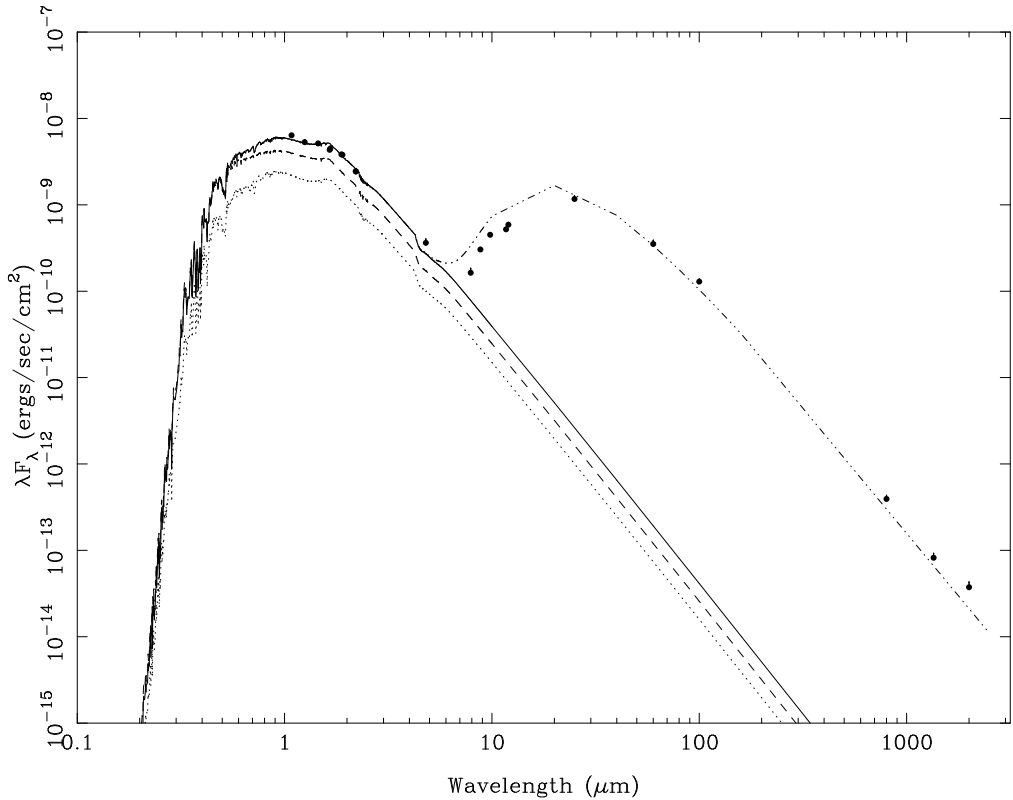}{0.1in}{270}{300}{380}{20}{0}
\caption{The stellar photosphere models for the components of
HD~98800~B (dashed = Ba, dotted = Bb, solid=total) along with
photometry from the literature.  The optical to near-infrared
photometry was used for the stellar fit, while the mid-infrared to
millimeter photometry was used for the disk modeling.  The
photometry is taken from \citet{pra01} and \citet{syl96} and where error
bars are not visible, they are smaller than the symbol size.  An example
flared disk model (dot-dashed) is shown with $T_o$=0.61,
$\Sigma=0.035$ gm/cm$^{2}$, $p=1$, $r_{in}=3.1$ AU, $\Delta r = 0.8$
AU and $\beta = 0.8$.
\label{fig:sed}
}
\end{figure}

All parameters have a range of values which are contained within
models which roughly match the SED.  Table 2 gives the 90\%
probability range for each parameter.  We note that these models tend
to produce too much flux in the mid-infrared, which can be eliminated
by removing the surface layer and having a single layer disk; however,
the probability distribution for the inner radius for the single layer
models does not include any models where $r_{in} > 3$~AU (fitting
the dynamical constraints). It may be possible to produce
a better fit to the mid-infrared SED and satisfy the dynamical 
constraints with modifications to the temperature structure of
the disk, but we did not wish to add more parameters. In
comparison to the single temperature fits \citep{low99, pra01} the
interior temperatures at the inner radius are below 150~K,
but this is balanced by the much hotter surface layer.  An example
model, along with the SED data used for the stellar photosphere fit
(\citet{bod05}) and the data used in the disk modeling, are shown in
Figure \ref{fig:sed}.  The SED can be roughly reproduced with a range
of inner radii from 1.1 to 3.6 AU. 
Interestingly, all the probable models have a relatively small radial
extent ($\Delta r < 2$~AU) with the 90\% probability range from 0.8 to
2.0 AU.  A disk with this extent would appear unresolved in the
mid-infrared imaging \citep[consistent with][]{koe00,pra01}.

We can now calculate if the models which reproduce the SED can also
explain the observed extinction.  Using an example model with $r_{in}
= 3.2$~AU, $\Delta r = 0.8$~AU and a surface density of 0.035
gm/cm$^2$ at a radial distance of 2 AU, the column density along the
derived line of sight is $\sim10^{-12}$ gm/cm$^2$, with similar column
densities derived for other models with high relative probabilities.
For the \citet{chi97} model visible dust opacity of $\kappa_V \sim
400~$cm$^2$/gm, this results in an extinction of $A_V \sim 10^{-11}$,
many orders of magnitude below the observed extinction of $0.3 \pm 0.05$.

\subsection{Vertically extended inner radius}

Next, we considered if the extinction could
arise from a vertically extended rim at the inner disk radius.
Vertically-extended inner rims arise from direct irradiation
of dust \citep{nat01,dul01} at the inner disk edge.   From \citet{dul01}
the pressure scale height is
\begin{equation}
h_{rim} = \frac{k T_{rim} R^{3}_{rim}}{\mu_g G M_{\star}}^{1/2}
\end{equation}
where $h_{rim}, T_{rim}$ and $R_{rim}$ are the pressure scale height,
temperature and radius of the inner rim, $\mu_g$ is the mean
molecular weight of the gas
and the rim height $H_{rim} \approx 5h_{rim}$.  Using a
combined mass of 1.28 M$_{\odot}$, $\mu_g=2$ and T$_{rim}$=150~K, 
\begin{equation}
H_{rim} \approx 0.12 \left(\frac{R_{rim}}{AU}\right)^{1/2} {\rm AU}.
\end{equation}
If the inner disk is dynamically truncated by the binary orbit at a radius of 3.4 AU (\S \ref{orbit}),
the height of the rim is 0.41 AU.  A rim at this radius and height
would obscure the line of sight starting at an inclination angle
of 85\degr.  Given the inclination angle of the binary orbit of 67\degr $\pm$ 3\degr,
the inner disk rim will not obscure the line-of-sight if the disk
is co-planar with the binary orbit.

\subsection{Circumbinary envelope}
\label{envelope}

Finally, we consider the possibility that the extinction towards \HD\
could arise from material distributed in a circumbinary envelope
around the B pair.  Assuming a spherical distribution, the surface
density calculated above can be used to estimate the total mass in
such an envelope if the inner and outer radius are known.  For the
purpose of a rough mass estimate, we use a value for the inner radius
from the region cleared by the B orbit of 3.4 AU and a value for the
outer radius of 10 AU (\S \ref{orbit}), which results in an envelope
mass of $5 \times 10^{-7}$ M$_{\odot}$.

Circumstellar envelopes are common around young stars with ages of a
few Myr but have generally dissipated at the derived age of \HD\
\citep[e.g.,][]{mun00}.  However, the dynamics of the quadruple system
are significantly more complicated than considered in the canonical
star formation picture and a small amount of material from the
formation of the central stars may have been unable to accrete onto
the disk.  We consider this unlikely given that the A component, which
has a similar mass and is presumably coeval has no such material.

\section{Dynamical simulations}
\label{dynamics}

The \HD\ circumbinary disk exists within the dynamical environment of
the larger quadruple system and could be warped by interactions with
HD~98800~A. To model the HD 98800 dynamical environment, we consider a
system comprising 3 stars.  Two of these stars represent the
HD~98800~B system with masses of $M_1 = 0.699 M_\odot$ and $M_2 =
0.582 M_\odot$, an orbital period of 315 days and an eccentricity of
$e = 0.78$.  These two stars are then surrounded by a circumbinary
disk comprising $25000$ test particles with an initial inner radius of $2$ AU
and an outer radius of $15$ AU.  The disk particles were distributed
randomly with no initial eccentricity, but with initial inclinations
that were Rayleigh-distributed, with the sine of the inclinations,
$\sin i$, having a root mean square value of $0.05$.  The disk is
therefore initially thin, but not completely flat.

The HD~98800~A binary system was then represented by a single star
with a mass equal to the combined mass of the two stars representing
the HD~98800~B system ($M_3 = 1.281 M_\odot$). Although we could have
modeled HD~98800~A using two stars, the actual orbital parameters and
stellar masses are not well known, and this would have been
computationally expensive.  Representing HD~98800~A with a single star
is an approximation that should at least give us some idea of how this
outer system influences the disk around HD~98800~B.  The orbital
period of the A-B system was taken to be $345$ years with an
eccentricity of $e = 0.5$ \citep[Orbit II]{tok99} and with an orbital
plane initially inclined at $10\degr$ to the plane of the disk around
HD~98800~B, which is coincident with HD~98800~B's orbital plane.

In this dynamical simulation we have ignored radiation pressure and
Poynting-Robertson drag \citep{burns79}.  HD~98800 is a young system
and the disk around HD~98800~B is relatively massive and is
collisionally dominated \citep{wyatt06}. Poynting-Robertson drag in a
system like HD~98800~B is therefore expected to be insignificant
because dust will be destroyed by collisions
before Poynting-Robertson drag can cause significant inward migration
\citep{wyatt05}.  Although radiation pressure could change the orbital
parameters of particles produced by collisions \citep{wyatt99},
the pericenters of these orbits should be randomly orientated and
so the effect will be to spread these particles out without changing
the plane of their orbits.  Since this is unlikely to affect
significantly the global structure of the disk we ignore it here and
include only the gravitational influence of the stars.

The system was evolved for just over $10^6$ years using a Hermite
integrator \citep{mak91}, that conserved both energy and angular
momentum to within $0.1 \%$. We found that the star
representing HD~98800~A caused the HD~98800~B binary system and its
circumstellar disk to undergo torque induced precession with a period
of $\sim 99000$ years.  The inclination of the HD 98800 B system, with
respect to the initial plane of the HD 98800 A/B orbit, also varies
sinusoidally from the initial 10$\arcdeg$ to just over 20$\arcdeg$
with a period of $\sim$43000 years.

\begin{figure}[!ht]
\includegraphics[scale=0.305]{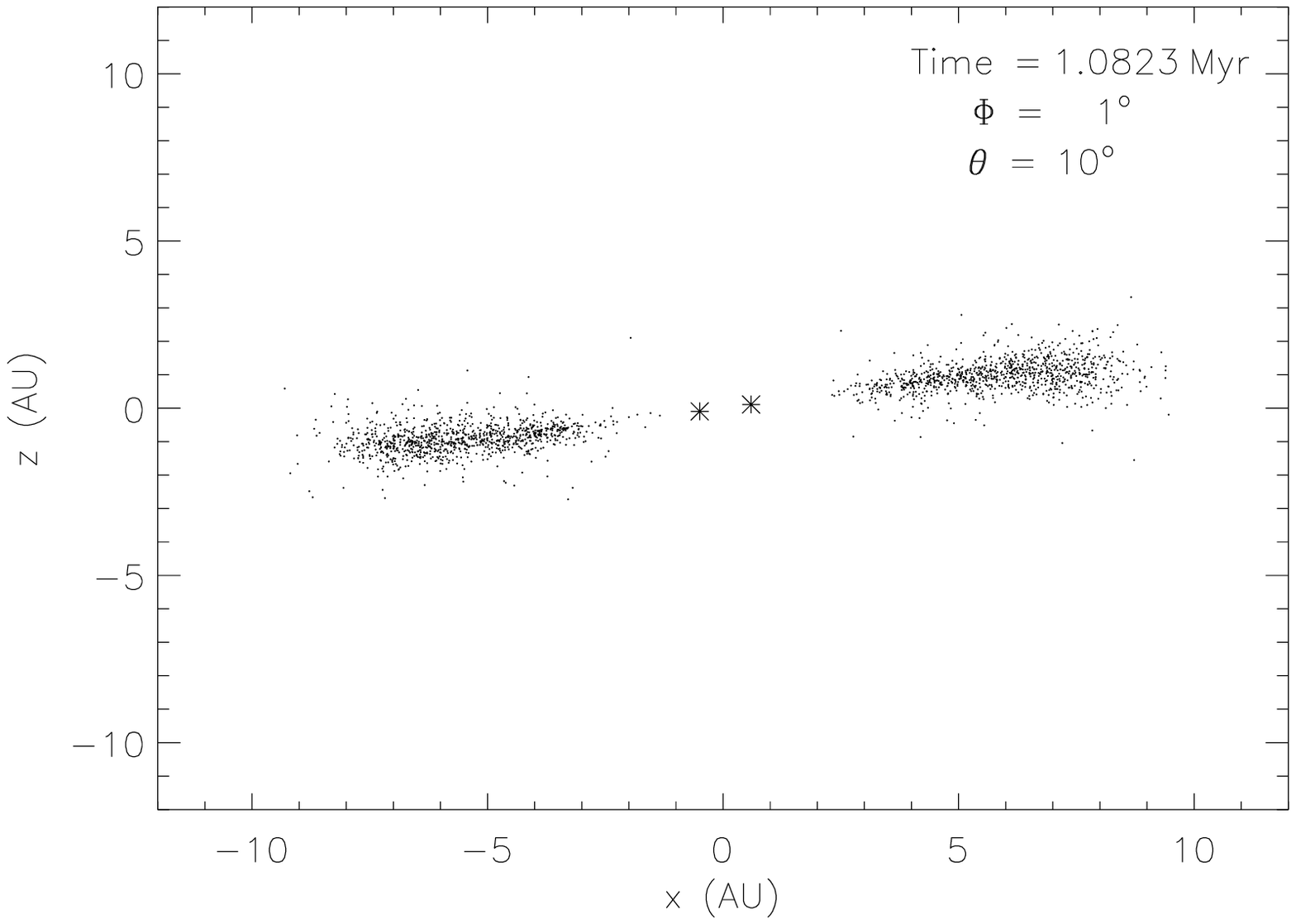}
\includegraphics[scale=0.305]{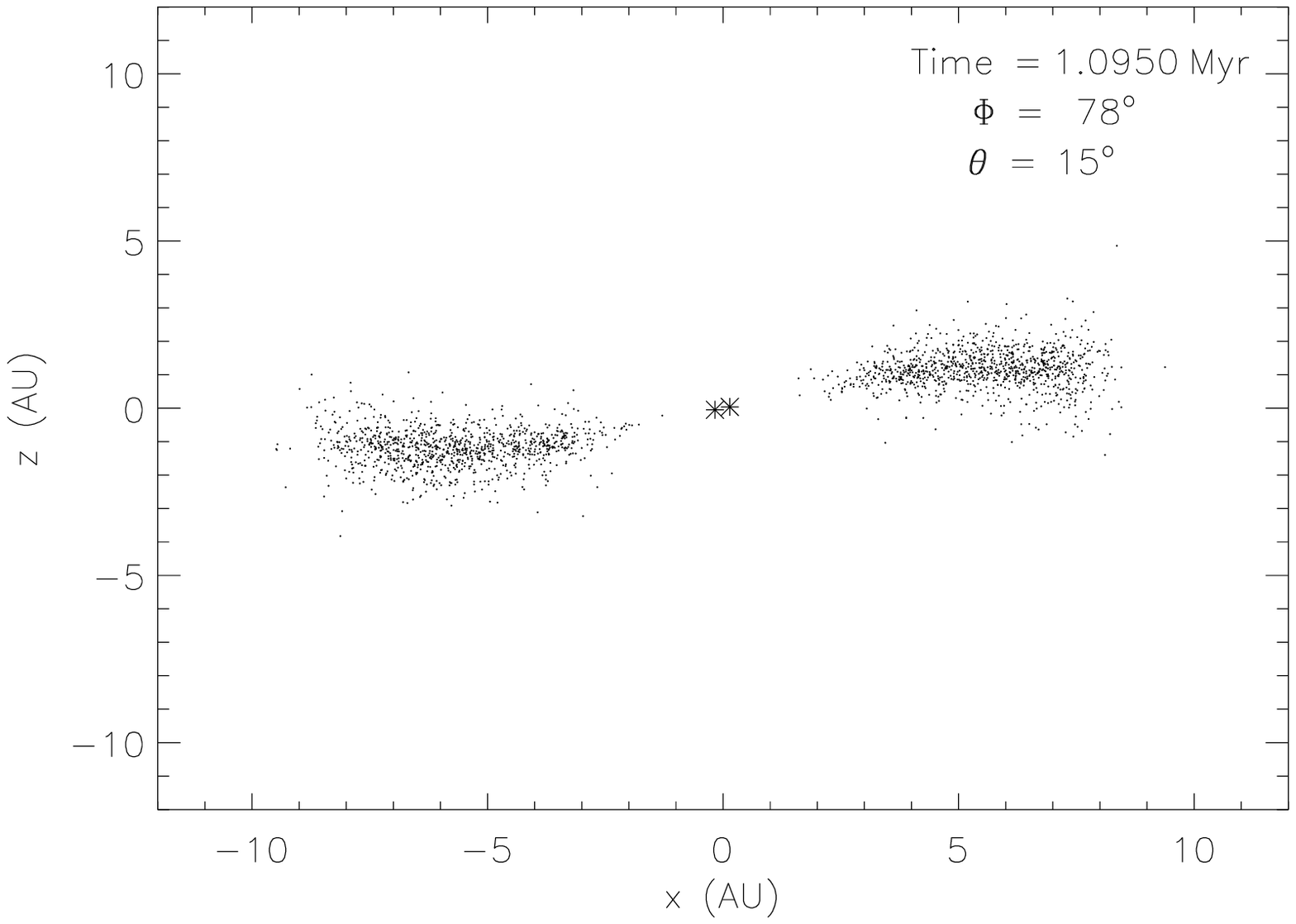}
\includegraphics[scale=0.305]{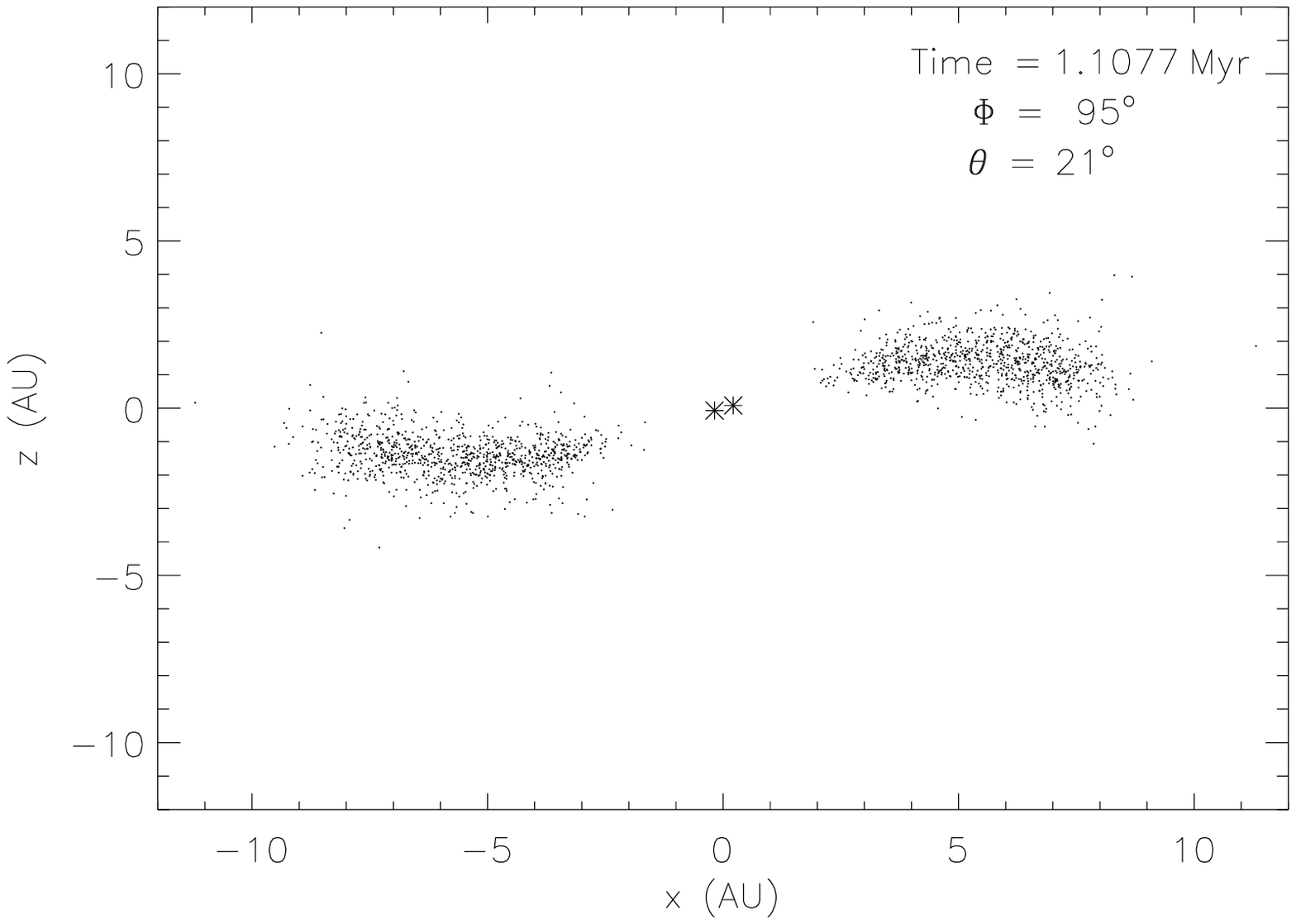}

\includegraphics[scale=0.305]{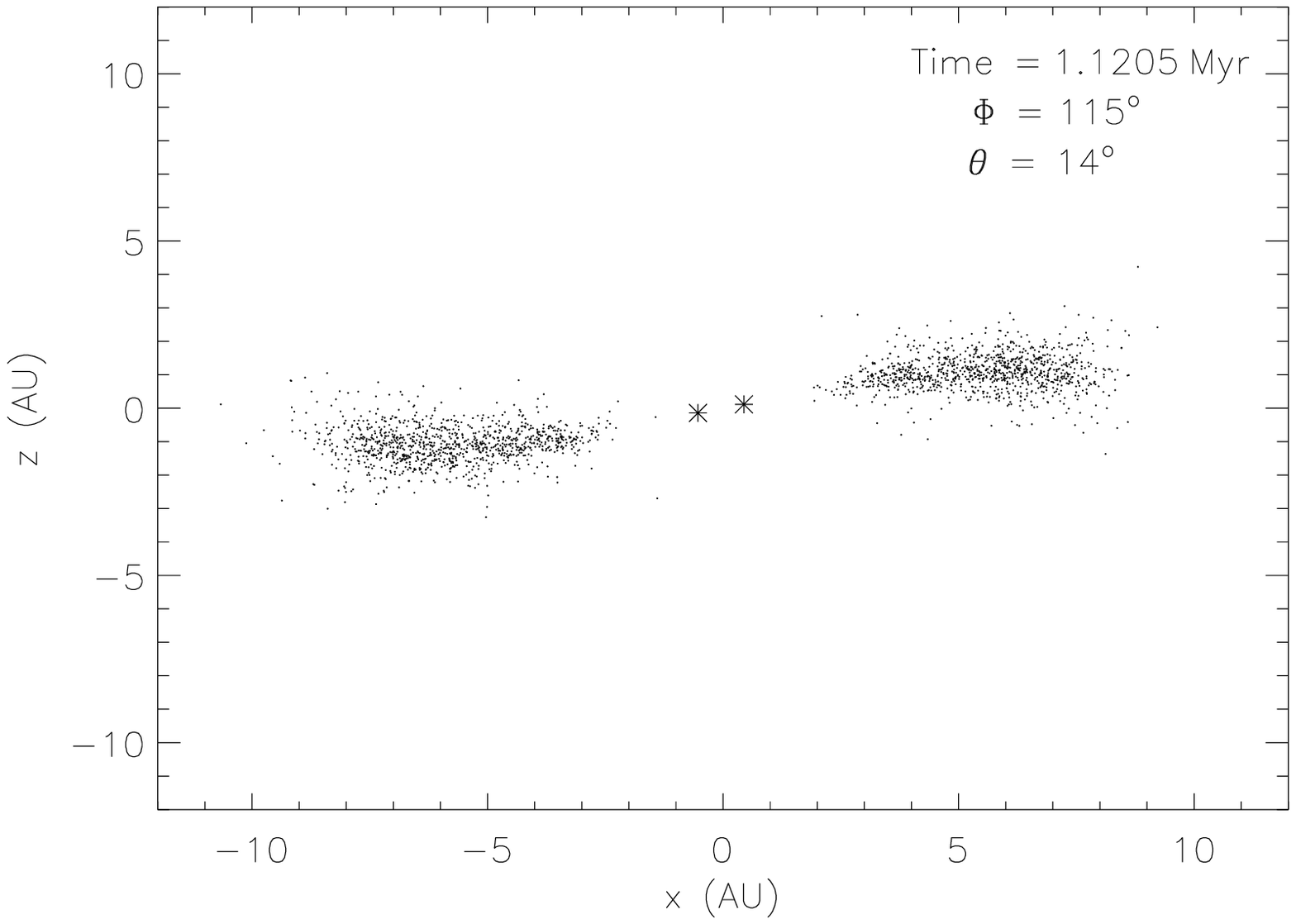}
\includegraphics[scale=0.305]{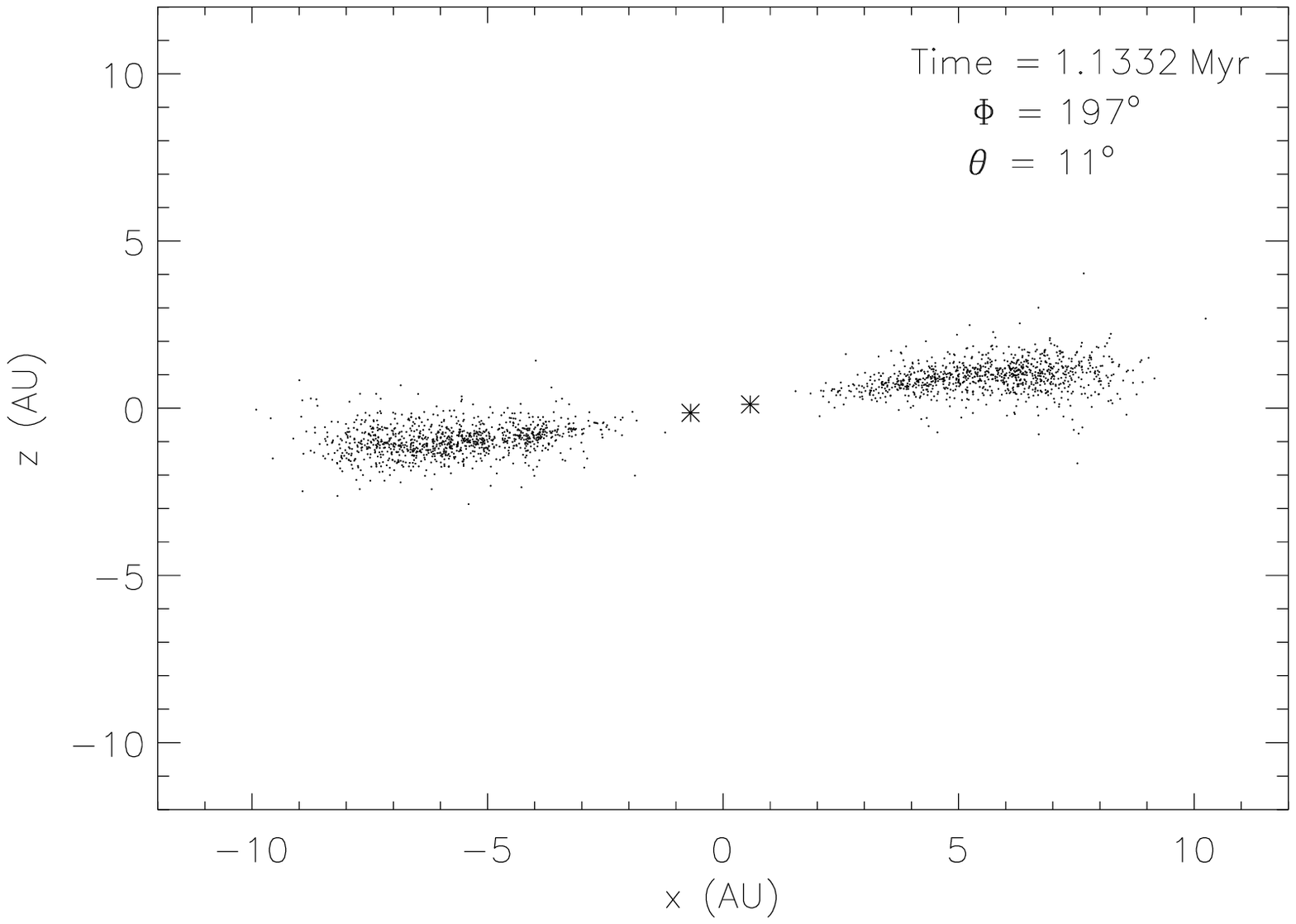}
\includegraphics[scale=0.305]{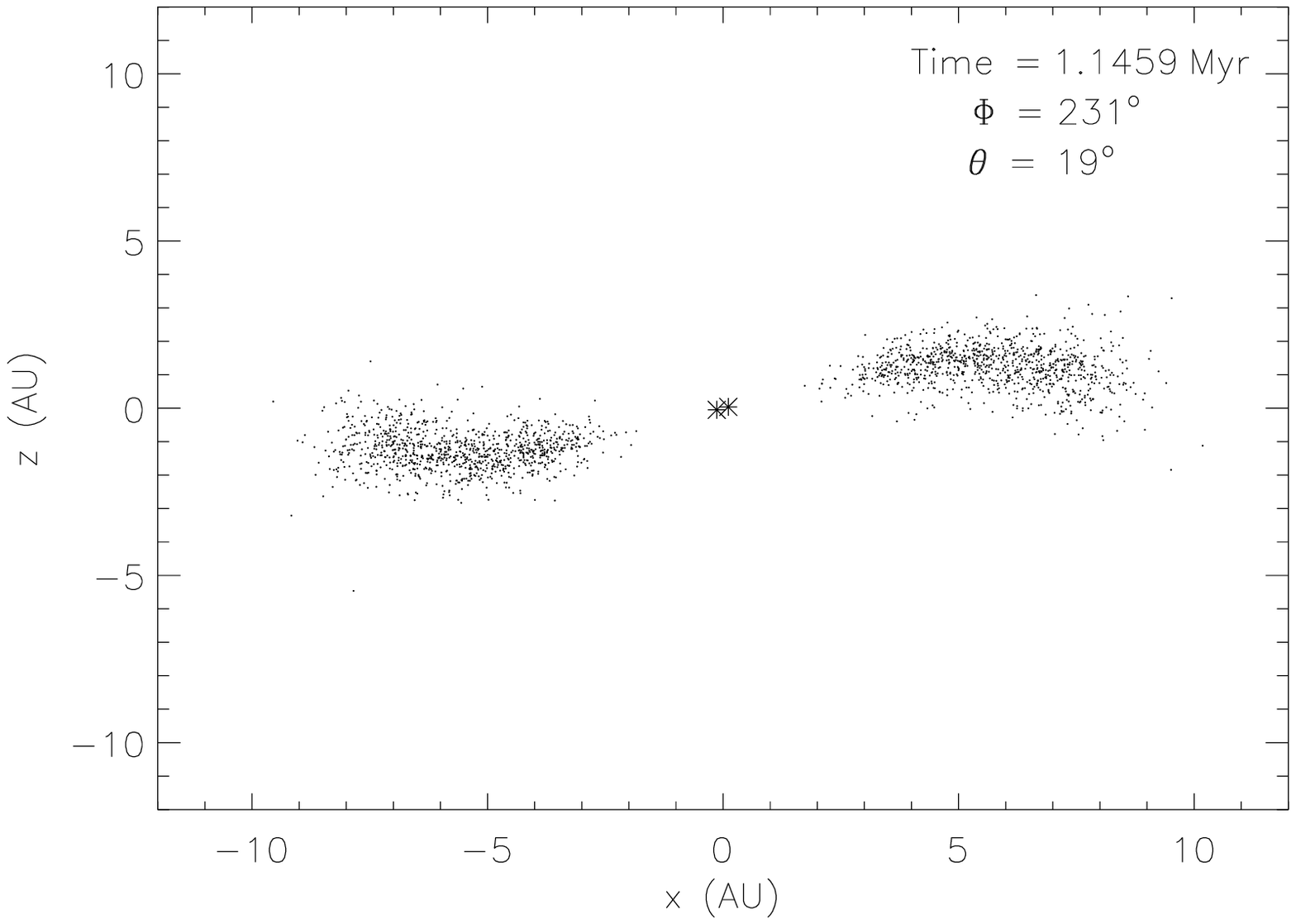}

\includegraphics[scale=0.305]{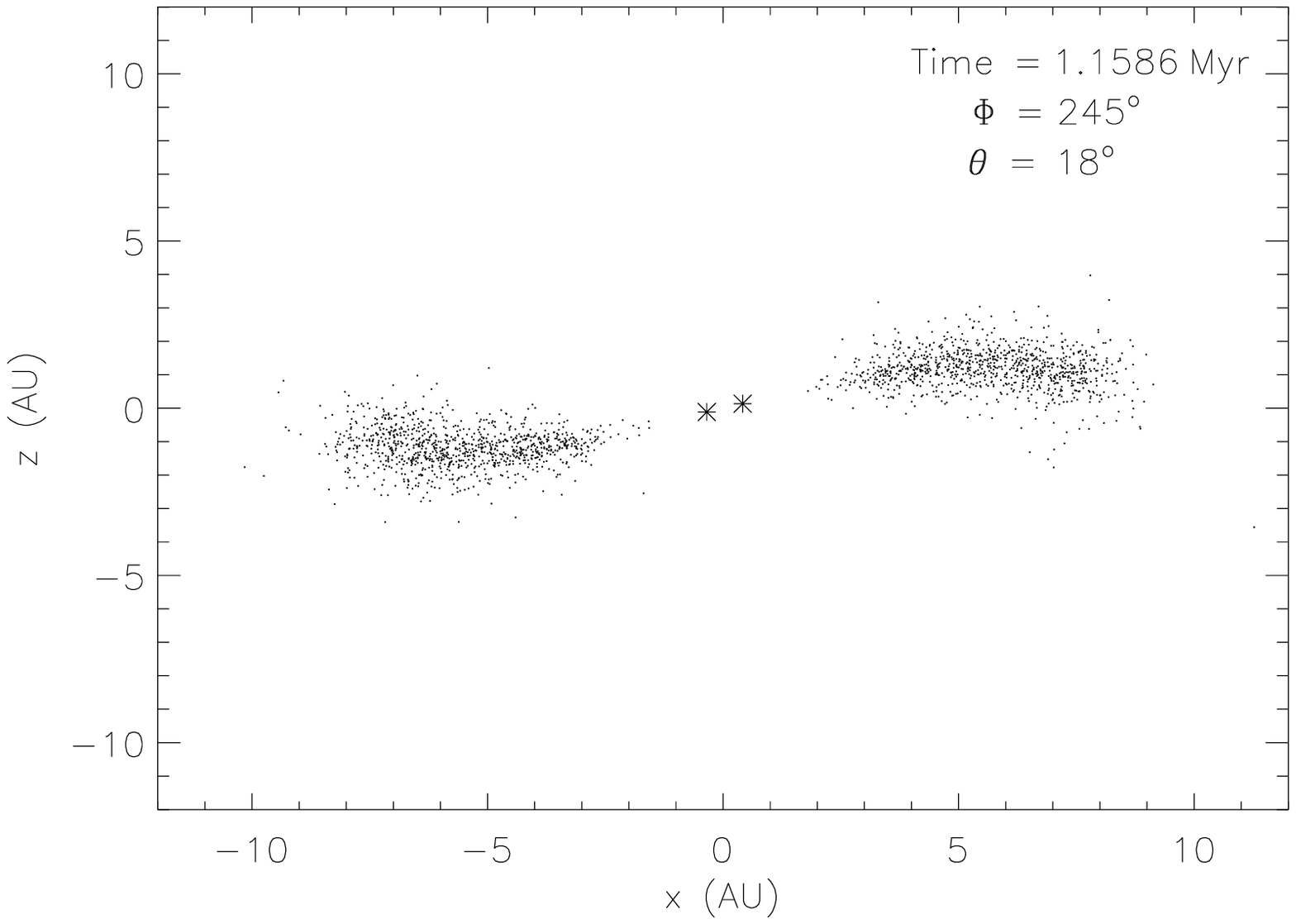}
\includegraphics[scale=0.305]{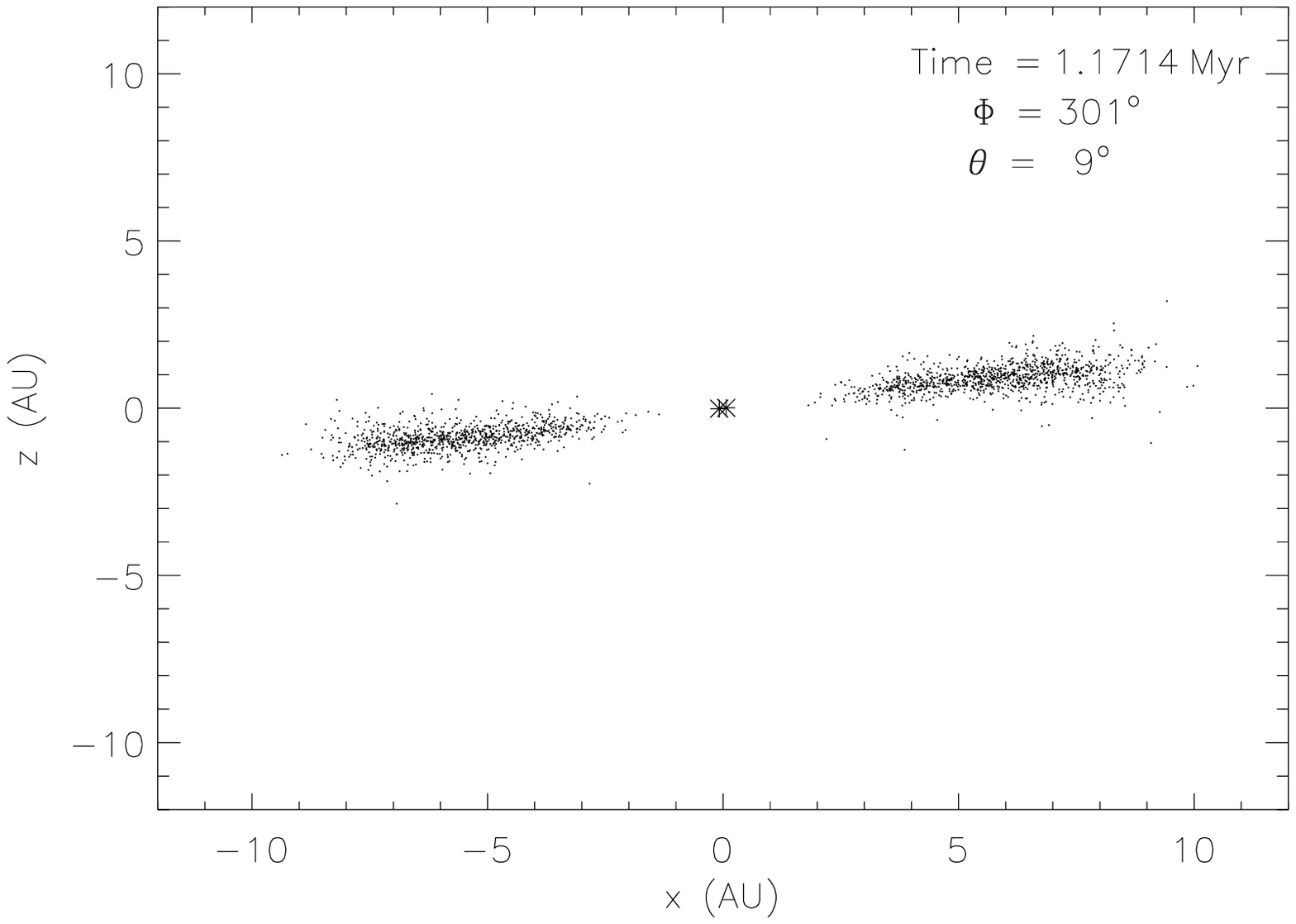}
\includegraphics[scale=0.305]{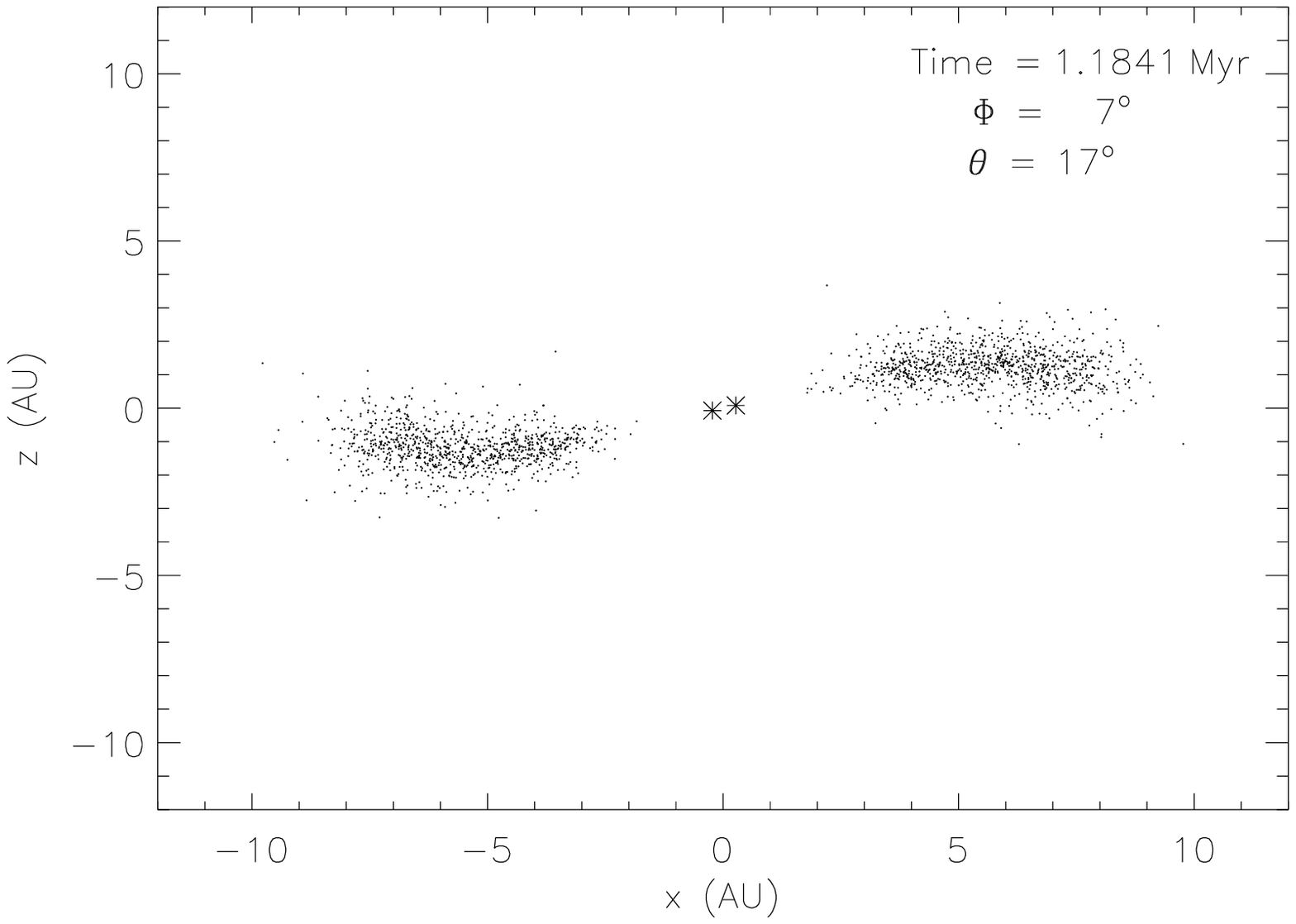}

\caption{A series of figures showing the evolution of the circumbinary
disk around HD~98800~B over one precession period. The center of mass
of HD~98800~B is located at the origin and the x-y plane is the
initial plane of the HD~98800~A/B orbit.  Each panel shows a slice
through the circumbinary disk together with the two central stars.
The viewing angle ($\phi$) is varied to take into account the precession of the
system and such that the disk is always viewed along the axis, the in
x-y plane, about which the disk appears to have been rotated.  Not
only do these figures illustrate how the inner and outer disk are
truncated, they also illustrate how the disk becomes warped as the
angle, $\theta$, between the orbital plane and the x-y axis increases.
This warping is qualitatively consistent with the visual extinction
observed towards HD~98800~B, which has an inclination angle of $67^o$.
\label{fig:warp}
}
\end{figure}

Figure \ref{fig:warp} shows the variation of the disk structure over
one precession period from $t = 1.0823$ Myr to $t = 1.1841$ Myr.  In
each panel the center of mass of the HD~98800~B system is located at
the origin, and each panel shows a thin slice through the circumbinary
disk together with the central stars of the HD~98800~B system.  In
these figures, the x-y plane is the initial plane of the orbit of the
HD~98800~A/B system.  The orbital plane of HD~98800~B and its
circumbinary disk were therefore initially inclined at $10\degr$ to
the x-y plane.  The label on each figure shows the time, the viewing
angle ($\phi$), and the inclination of the orbital plane of HD~98800~B
relative to the x-y plane ($\theta$).  The viewing angle is varied to
take into account the precession of HD~98800~B, such that the disk is
always viewed along the axis, in the x-y plane, about which the disk
appears to have been rotated.

What the panels in Figure \ref{fig:warp} show is that the eccentric
orbit of HD~98800~B ($e = 0.781$) truncates the inner disk at a radius
of $\sim 3$ AU (in agreement with the estimates in \S \ref{orbit}),
while the outer disk, which initially extended to a radius of 15~AU, is now
truncated at a radius of $\sim 10$ AU due to the influence of
HD~98800~A.  Note that this outer truncation radius also agrees with
the estimates based on co-planar interactions (\S \ref{orbit}).  This
outer radius value further supports the idea that the observed visual
extinction to HD~98800~B is unlikely to be due to a flared co-planar
disk extending to large radii.  What is also clear from Figure
\ref{fig:warp} is that when the inclination of the orbital plane
relative to x-y plane is small ($\theta \sim 10\degr$), the disk is
thin and slightly flared.  However, as the angle increases, the disk
becomes more obviously warped.  Consider the third panel in the top
row. The orbital plane is inclined at $21^o$ to the x-y axis and yet a
line of sight along the x-axis would still intercept some disk
material.

Although the resolution of our simulation is insufficient to quantify
the column density of material located more than $20\degr$ above the
orbital plane of HD~98800~B, the existence of the warp is
qualitatively consistent with the observed extinction to HD~98800~B,
the orbit of which has an inclination angle of $67\degr$, particularly
given the low column density required to produce the observed
extinction.  This is consistent with the observed variability in the
Hipparcos data \citep{sod98} arising from changes in extinction if the
disk material is clumpy, but as these simulations can not quantify the
column density, we can not directly compare to the variability.  The
rotation time scale of the warp is much longer than the 14 day
periodic variations observed by \citet{hen94} which does not correspond
well to the disk rotation time scale of 2.5 yrs at 3 AU.

\section{Discussion}
\label{discussion}

In this section, we first compare our results on the circumbinary
disk of \HD\ to models of this disk from other authors.  Next we
compare the \HD\ disk to samples of other objects with primordial and
secondary disks, discuss the implications for the evolutionary status
of this object and suggest some observational tests of this status.

\subsection{Comparison to other models}

As listed in \S \ref{intro} and \S \ref{disk}, several other groups
have modeled the disk of \HD\ \citep{low99,koe00,pra01,fur07}.  All
these models have the following characteristics in common: a
circumbinary disk with a inner region cleared to at least a few AU
dominated by emission from material with a temperature of $\sim$150~K.
Here we compare to the \citet{pra01} and \citet{fur07} models which
use either imaging constraints or detailed spectroscopy in their
modeling.  Using their unresolved images as constraints,
\citet{pra01} fit the SED with a blackbody temperature of 150~K and
derived an inner radius 2 AU, an outer radius of at least 5 AU, and a
possible cooler dust component at larger radii which is then truncated
by the A/B component interactions.  They also comment that the
vertical extent must be $\sim$1.5~AU to account for the high
fractional luminosity.  Our dynamical simulations and the work of
\citet{art94} and \citet{pic05} suggest that this inner radius is too
small to be stable given the orbital parameters of the binary.

Recently, \citet{fur07} used the mid-infrared spectrum of HD~98800
from the {\it}Spitzer telescope to constrain the structure and
composition of the \HD\ disk.  Their spectrum shows no excess below
5.5 $\mu$m (within a 5\% absolute calibration) and a strong excess
starting at 8 $\mu$m.  They use the structure in the silicate feature
to model the dust composition and we refer interested readers to the
paper for more details on that aspect.  In comparison to our model,
the relevant model parameters are an optically thick region starting
at 5.9 AU with a temperature of 150~K and an optically thin ring
between 1.5 and 2 AU with grain sizes of a few microns and a
temperature of 310~K where the silicate emission features arise from.
This inner ring is within the apastron separation of the Ba-Bb pair of
1.75 AU and within the dynamically cleared regions in the work of
\citet{art94}, \citet{pic05} and \S \ref{dynamics}.  If material is
located at these radii, it will be transient as the time scales for
clearing the gap ($<$100 orbital periods, i.e.  $<$100 yrs) are much shorter
than the P-R drag timescale of 10$^5$~yrs for a 3 $\mu$m grain at 3 AU
in this system.  \HD\ also shows no indication of accretion, unlike
the DQ~Tau and UZ~Tau~E systems which are inferred to have active
accretion through streamers from a circumbinary disk
\citep{mat97,jen07}.

Both \citet{pra01} and \citet{fur07} noted that small grains must be
present in the \HD\ disk to account for the spectral features.
A distribution of grains size is inferred in many pre-main
sequence, primordial disks \citep[see review by][]{nat07} and the
higher temperatures silicate grains inferred by \citet{fur07} from
features in the mid-infrared spectrum could reside on the stellar
photon-illuminated surface of a flared or warped disk such as that
discussed in \S \ref{flared}.  In the example model presented in
Figure 2, the interior temperature at the inner rim is 60~K, while the
surface layer temperature is 210~K.   Spectrally
resolved mid-infrared interferometry observations would constrain the
location of the small grains responsible for the silicate emission
feature.

\subsection{Comparison to other primordial and secondary disks}

The TW Hya association is unusual in containing sources with a very
large range of pre-main sequence properties. Four of the sources have
{\it IRAS}-detected infrared excesses (TW~Hya, HD 98800, Hen 3-600 and
HR 4796A).  TW Hya and Hen 3-600 are often classified as T Tauri stars
given their accretion signatures \citep{muz00} and near-infrared
excess, and TW Hya has observed CO and H$_2$ in the disk
\citep{den05,wei00}.  On the other hand, HR 4796A has a similar SED to
HD 98800, but with material at much larger radii and is consistent
with a secondary dust debris disk.  Thus, it is not clear from the
association as a whole whether the \HD\ disk is primordial or
secondary.  Here we compare the derived disk properties of \HD\ with
those of other primordial/protostellar and secondary/debris disk
systems.

The mass distribution of the most probable models from \S \ref{flared}
for a flared gas and dust disk peaks at $10^{-7}$~M$_{\odot}$,
significantly lower than most classical T Tauri stars and comparable
to many secondary debris disk stars such as the beta Pic moving group
(e.g.  \citet{mey07}); however, as shown in \S \ref{dynamics}, the
\HD\ circumbinary disk is likely to be dynamically truncated at both
the inner and outer radii, and the disk mass is highly correlated with
the small radial extent (few AU) as compared to values of $\sim$100 AU
typical for T Tauri circumstellar disks.  An additional contribution
to the low disk mass is the surface density of $<$0.5 gm/cm$^2$ at a
radius of 2 AU, which is a factor of $\sim$100 lower than the minimum
mass solar nebula.  The high infrared luminosity of the disk (0.1 --
0.2 L$_{\ast}$) has been noted by many groups
\citep{zuc93,koe00,pra01} and this fraction is significantly higher
other young debris disks systems, such as the sample observed by
\citet{che06} of A to M spectral types with ages $<$50~Myr.  In that
sample, the detected systems had infrared disk luminosities of $5
\times 10^{-6} - 5 \times 10^{-3}$ L$_{IR}$/L$_{\ast}$.  However, the
dust in those systems is in general located many to tens of AU from
the central star, rather than a few AU and is not likely to be as
vertically extended.  Overall, the disk properties for \HD\ are
intermediate between those of T Tauri, primordial disks (and its
association members TW Hya and Hen 3-600) and secondary debris disks
around stars with ages of tens of Myr.

A more stringent diagnostic of primordial vs. secondary material would
be to observe gas tracers in the disk and establish the gas to dust
ratio.  In a CO survey by \citet{den05}, HD~98800 was not detected
with an upper limit of 0.09 K~km/sec, while TW Hya was detected with
2.16 K~km/sec of emission.  Although their relative infrared excesses
($\sim$0.1 for HD 98800 and 0.3 for TW Hya) are similar, the
difference in CO emission strengths may result primarily from the CO
arising in colder gas 10's of AU from the star.  As the \HD\ disk is
probably much smaller than this, the CO non-detection does not set
stringent limits on gas depletion.

The lack of accretion signature in \HD\ is not surprising in the
context of a $\sim$3~AU inner hole; however, accretion across the gap in
a circumbinary disk can take place \citep{art96}, particularly for
geometrically thick disks and has been observed in T Tauri
circumbinary disks such as DQ Tau and UZ~Tau~E \citep{mat97,jen07}.
Dynamical modeling by \citet{art96} shows that the mass accretion
from circumbinary disks is strongly modulated in time and peaks at
periastron.  Thus another observational test for the \HD\ disk would
be time monitoring of accretion diagnostics, particularly centered
about periastron.

\section{Summary}
\label{summary}

\HD\ is one of the few young multiple stars with significant disk
emission for which the physical orbit has been determined.  We have
used the orbital parameters, particularly the inclination, and the
assumption of disk-binary coplanarity to examine physical models for
the disk.  As first discussed by \citet{pra01} the large mid-infrared
excess and unresolved images of \HD\ seem to require an optically and
geometrically thick disk.  Either a flared disk or a vertically
extended inner rim can satisfy the geometric requirement, but
dynamical truncation will keep the majority of the material beyond an
inner rim with a radius of 3~AU.  The disk solution presented here is
not unique in reproducing the SED and somewhat over-produces mid-infrared
continuum flux, but does satisfy the dynamical constraints.

We find that a standard disk prescription is unable to explain the
observed extinction if the disk and the binary orbit are co-planar.
For the flared disk model in Figure \ref{fig:sed} to intercept enough
of the line of sight to produce the observed extinction, the disk
inclination would need to be 81$\arcdeg$ as compared to the binary
orbit inclination of 67\degr.  From the dynamical simulations shown in
\S \ref{dynamics}, interactions between the B circumbinary disk and
the A component are sufficient to perturb the disk inclination by that
amount.  They are also sufficient to warp a thin disk enough to place
material into the observed line of sight.  In either case, we suggest
that the small observed extinction is due to significant perturbations
of the circumbinary disk by the A binary component in this quadruple
system.  The next step in understanding the \HD\ disk would
be to quantitatively combine the SED and observational imaging constraints
(unresolved mid-infrared size and extinction along the line of sight)
with the dynamical interactions between all the components in the
HD~98800 system.

\acknowledgments

We thank Sverre Aarseth for help with the use of his N-body codes and
Kenneth Wood for initial SED modeling.  We also thank Peter Brand and
Mark Wyatt for useful discussions about secular perturbation theory.
We are grateful to the anonymous referee for several helpful suggestions.

\end{document}